# Cosine bands, flat bands and superconductivity in orthorhombic iron selenide

*Ian D. R. Mackinnon[1]\* and Jose A. Alarco[2]*

Ian D. R. Mackinnon

[1]School of Earth and Atmospheric Sciences, Queensland University of Technology, Brisbane, QLD 4001, Australia

Jose A. Alarco

[2]School of Chemistry and Physics, and Centre for Materials Science, Queensland University of Technology, Brisbane, QLD 4001, Australia

Email: ian.mackinnon@qut.edu.au

[1]ORCID: 0002-0732-8987; [2]ORCID: 0001-6345-071X;

Funding: This research received no external funding

Electronic band structures (EBSs) for orthorhombic beta–$FeSe_{1-x}$ at <16 K and up to 23 GPa using experimentally determined cell dimensions are evaluated for cosine-shaped bands near, or crossing, $E_F$. Cosine-shaped bands are present in reciprocal directions parallel to the *c* axis at all pressures. Calculations using a P1 cell derived from Cmma symmetry with a 2*c* superlattice moderates the effect of intersecting bands to ~9.0 GPa. This approach enables determination of a superconducting gap consistent with experimentally determined values. Key influences on charge distribution and transfer in the interplanar region of beta–$FeSe_{1-x}$ are lone pair electrons which feature as flat bands (FBs) near $E_F$ along G–Z in an EBS. FBs also influence the topology of Fermi surfaces as pressure increases and in directions parallel to the $c^*$ direction (i.e. offset along $k_y$) within the Brillouin zone. At the Fermi surface along $b^*$, cosine bands split and align favourably for electron-hole pairing with nodal inflection points located at $E_F$. For P ≥ 12.0 GPa, FBs interact with folded cosine bands invoking additional band dispersions. These calculations suggest that FBs participate in, and with increased pressure, enhance and sustain the superconducting properties of beta–$FeSe_{1-x}$ to ~23 GPa.
.

1. Introduction

Since the first observation of superconductivity in iron selenide ($FeSe_{1-x}$) at ambient pressure[1] early investigations focused on structural transformation(s) with temperature[2-5] and pressure.[2, 6-9] At ambient pressure, tetragonal β–$FeSe_{1-x}$ (SG: P4/nmm) undergoes a structural transition at ~90 K and completes transformation to a discernible orthorhombic structure (SG: Cmma) using diffraction techniques at ~70 K.[3] The temperature range for this transition to orthorhombic Cmma is dependent on the stoichiometry or disorder of β–$FeSe_{1-x}$.[10] The superconducting transition temperature, $T_c$, for polycrystalline β–$FeSe_{1-x}$ is ~8.5 K at ambient pressure for $0.0 < x \leq 0.02$ (i.e. a Se deficiency of ~0.020 ± 0.005).[8, 11-13]

Successful growth of phase pure mm-sized single crystals[14, 15] of β–$FeSe_{1-x}$ narrowed the stoichiometry for maximum $T_c$ of the orthorhombic form to $0.000 < x < 0.010(5)$ at ambient pressure.[15] Syntheses of polycrystalline β–$FeSe_{1-x}$ show that while stoichiometry is important for achievement of maximum $T_c$, variations to chemistry, processing and the presence of other phases (e.g. α–Fe, δ–$Fe_xSe$) does not extinguish superconductivity.[12, 16-20] Nevertheless, magnetization studies on polycrystalline β–$FeSe_{1-x}$ at ambient pressure show that superconductivity is suppressed with large deviations from stoichiometry, particularly as Se vacancy content increases[5, 11] or with metal site substitutions for Fe.[21]

Single crystal experiments, recently summarised by Xu and Xie,[22] show that other influences such as order-disorder,[10, 23] carrier density[24], or strain[24] can affect the value of $T_c$ for β–$FeSe_{1-x}$. Ma et al.[24] detail these influences as well as the effects of processing conditions (see Tables S8 and S9 of [24]). At or near stoichiometric composition (e.g. $0.0 < x < 0.02$), the presence of Fe as interstitials in β–$FeSe_{1-x}$, or the presence of other minor phases such as α–Fe, δ–$Fe_xSe$ and γ–$Fe_xSe$ appear to have limited impact on the value of $T_c$ (e.g. ± 1.5 K) for β–$FeSe_{1-x}$ at low or ambient pressure.

The study by Medvedev et al.[8] used a phase pure powder sample of $FeSe_{1-x}$ with x = 0.01 to track the $T_c$ with applied pressure. With increase in applied pressure, the value of $T_c$ for β–$FeSe_{1-x}$ increases to an experimentally determined maximum of 36.7 K at 8.9 GPa and decreases with higher applied pressure.[8] This pressure-$T_c$ effect is counter to that for the well-known layered superconductor, $MgB_2$.[25] Systematic crystallographic studies at low temperature (i.e. < 70 K) with applied pressure show that orthorhombic β–$FeSe_{1-x}$ (SG: Cmma) is the dominant phase up to ~9.0 GPa.[6, 7, 9]

Extensive investigations of iron-based superconductors (FeSCs) have shown that electron-electron interactions and orbital differentiation of Fe orbitals are important features that distinguish these compounds from conventional electron-phonon coupled superconductors.[26]

A recent review[26] provides a comprehensive summary of experimental and theoretical frameworks that highlight the unique character of electron behaviour in FeSCs. For β–FeSe$_{1-x}$, these frameworks have focused on the Fe–Fe layer in the *x-y* plane of a well-defined two layer compound

DFT models of layered superconductors such as MgB$_2$, CaC$_6$ and LaH$_{10}$ show that electronic band structures (EBSs) contain features consistent with experimentally determined superconducting gap values.[27-29]. In this work, we calculate EBSs for orthorhombic β–FeSe$_{1-x}$ focussing on energies near the Fermi level (E$_F$) to evaluate electron distributions and bonding influences with change in pressure. To elaborate the detail in an EBS, as well as to identify a superconducting gap, cosine-shaped bands aligned with the *c* axis that are pinned to nodal points and are near to, or cross, E$_F$ are suitable candidates for examination.[29] A motivation is to use experimentally determined crystallographic data with *ab initio* DFT, to calculate the influence of pressure on electron/quasi-particle distribution, bonding and superconductivity.

2. Results

In this work, the electronic band structure (EBS) for orthorhombic β–FeSe$_{1-x}$ with change in pressure using experimentally determined structural data collected by Kumar et al.[7] at 8 K for x = 0.01 are evaluated. Structural data collected by Margadonna et al.[3, 6] at 5 K and 16 K on orthorhombic β–FeSe$_{1-x}$ (x = 0.03) are also evaluated. These experimental studies[3, 6, 7] detail the proportion of minor phases in the powder and present well-defined, refined cell parameters for each phase.

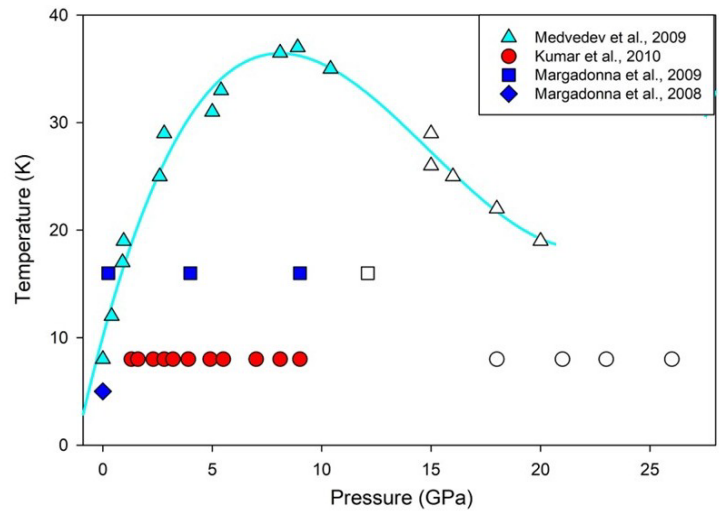

**Figure 1:** Experimental data for β–FeSe$_{1-x}$ showing (a) superconducting transition temperature (T$_c$) with applied pressure (triangles) after Medvedev et al.;[8] (b) Synchrotron X-ray and neutron diffraction data at 5 K (solid diamond),[3] 8 K (solid circles)[7] and 16 K (solid squares)

[6], respectively. Open symbols designate conditions for which orthorhombic β–FeSe$_{1-x}$ (SG: Cmma) is in low abundance in powder samples.

**Figure 1** shows the temperature-pressure (P–T) distribution of structural data points from these studies[3, 6, 7] in relation to experimentally determined values for T$_c$ with pressure.[8] DFT models for β–FeSe$_{1-x}$ are linked to real space structural data (at T, P shown in Figure 1) with electronic behaviour that influences bonding environments. Examples that illustrate structural or electronic features at key pressure(s) are shown in this section. Additional examples covering pressures up to 23.0 GPa are detailed in Supporting Information.

2.1. β–FeSe$_{1-x}$ Structures

Structures of β–FeSe$_{1-x}$ with dependent symmetry transformations are shown in **Figure 2**. These structures encompass the room temperature tetragonal β–FeSe$_{1-x}$ (SG: P4/nmm) which persists to ~90 K, orthorhombic β–FeSe$_{1-x}$ (SG: Cmma) and a primitive sub-lattice derived from the orthorhombic structure (SG: P1; see 5.0 Methods). The tetragonal and orthorhombic structures shown in Figure 2 are constructed with experimental data by Kumar et al.[7] and are used in DFT calculations shown below. The structural transformation of β–FeSe$_{1-x}$ with change in temperature to < 70 K is described in Supporting Information (**Figure S1**). This structural change involves a 45° rotation around the *c* axis of the tetragonal form at T < 70 K. The transformation matrix for conversion of a Cmma lattice to a P1 lattice (and vice versa) is also shown in Figure S1. Conversion of a tetragonal P4/nmm lattice to a P1 lattice[30] is not equivalent.

Distortion of the tetragonal lattice with pressure only (i.e. at ambient temperature) and with reduction in temperature (at ambient pressure), as well as to the orthorhombic lattice with reduced temperature and increased pressure are well documented.[3, 4, 6-9, 13, 17, 19, 31] Variations to unit cell parameters with change in pressure for the tetragonal and orthorhombic lattices are shown in Supporting Information (**Figure S2** and **Figure S3**; **Table S1**, and **Table S2**). The shortest interlayer Se–Se distances and key angles between atoms for the tetragonal and orthorhombic structures are listed in Supporting Materials (Table S2 and **Table S3**).

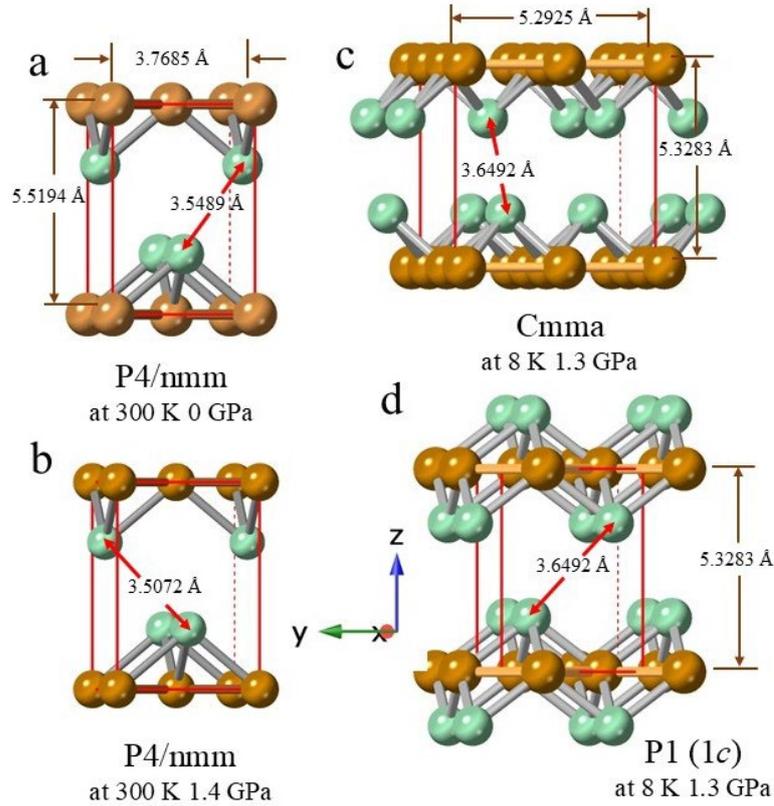

**Figure 2:** Crystal structure schematics showing space group symmetries used for DFT calculations. Structures are shown along [100] with a 10° rotation about the *c* axis to highlight all atoms in a unit cell. The interlayer Se–Se distances, based on experimental data,[7] are shown for (a) tetragonal β–FeSe$_{1-x}$ at 300 K and 0 GPa and (b) at 300 K and 1.4 GPa, (c) orthorhombic β–FeSe$_{1-x}$ at 8 K and 1.3 GPa and (d) a P1 1*c* lattice derived from orthorhombic β–FeSe$_{1-x}$ at 8 K and 1.3 GPa. Unit cells are outlined in red.

In general, substantial changes to interlayer distances, of magnitude between ~9% and 11% along the *c* axis direction, are evident for both tetragonal and orthorhombic forms between 0 GPa and 11 GPa. Orthorhombic β–FeSe$_{1-x}$ responds to applied pressure by adjusting atom-atom distances in a manner different to that by tetragonal β–FeSe$_{1-x}$. Structural response to applied pressure for orthorhombic β–FeSe$_{1-x}$ is illustrated by the variation of *a* and *b* cell dimensions with pressure as shown in Figure S2a (Supporting Information). The *a* axis alternates from shorter than, (e.g. at 1.3 GPa and 3.2 GPa) to longer than, the *b* axis (e.g. at 1.6 GPa, 2.3 GPa and 2.8 GPa) with applied pressure. These variations in shortest atom-atom distances are indicators of electron behaviour in β–FeSe$_{1-x}$ with pressure. In summary, the tetragonal and orthorhombic structures of β–FeSe$_{1-x}$ are distinctly different and warrant DFT calculations that reflect this difference. Electron distributions within each structure may look similar in an EBS, but result in dissimilar bonding configurations and physical properties.

2.2. Electronic Band Structures

A range of DFT calculations have been undertaken on β–FeSe$_{1-x}$ for the symmetry conditions shown in Figure 2. Details on these calculations are provided in Methods and in earlier work.[27, 32] DFT calculations use both LDA and GGA functionals in order to provide reliable error estimates of energy values for key features on either side of E$_F$.[33, 34] For derived parameters, data from each calculation (e.g. with GGA or LDA) are averaged.

2.2.1. Tetragonal and Orthorhombic Structures

Band intersection, band overlap or band crossing avoidance are often encountered in the EBS of high symmetry structures and can make interpretation of electronic bands problematic.[29] Band interactions are evident in the EBS for tetragonal and orthorhombic β–FeSe$_{1-x}$ as shown in Supporting Information (**Figure S6** and **Figure S7**). The reciprocal space projection along the $c^*$ axis, Γ–Z, is of interest as this provides useful information on the interplanar direction for both structure types.

Inspection of Figure S6 and Figure S7 shows that the EBS for a tetragonal structure is different to that of the orthorhombic structure when calculated using an LDA or GGA functional. Flat bands (FBs) are evident in the EBS for both structures and the Γ–Z section shows a band that appears to have cosine-like format. Similar band format is apparent on closer inspection of the Γ–Z section for both structure types (e.g. Figure S6c, S6d and Figure S7c) but show subtle differences in band characteristics. For example, the cosine-shaped curves in Figure S6 are intersected by two other crossing bands including a flat band, with in one case, apparent band avoidance and not in the other. Band crossings can obfuscate interpretation of an EBS, particulalry if instructive detail is sought from derived parameters (as detailed in Section 2.2.2). Earlier work shows that use of reduced symmetry for DFT calculations provides a facile means to interpret complex electron-phonon and electron-electron interactions of high symmetry structures.[27, 29, 34, 35]

2.2.2. Superlattice band calculations

For this study, orthorhombic β–FeSe$_{1-x}$ structures are converted to a P1 lattice using the transformation shown in Figure S1. EBS results for a 2$c$ superlattice of the P1 structure with geometry optimisation is equivalent to folding the EBS results of the P1 1$c$ structure half-way along the Γ–Z direction.

**Figure 3** shows the EBS for a 2*c* primitive lattice of β–FeSe$_{1-x}$ for two different stoichiometries and at different temperatures and pressures (at 5 K and 0 GPa compared to 8 K and 3.9 GPa) to highlight band energy changes. At 5 K (Figure 3a), the FB is below E$_F$, while at 8 K (Figure 3b) the FB is above E$_F$ and the energy difference between bands at the Γ node (circled) increases with increased applied pressure. The effects on an EBS with and without geometry optimisation, which substantially affect detail in the Z–Γ orientation, are shown in **Figure S7** (Supporting Information).

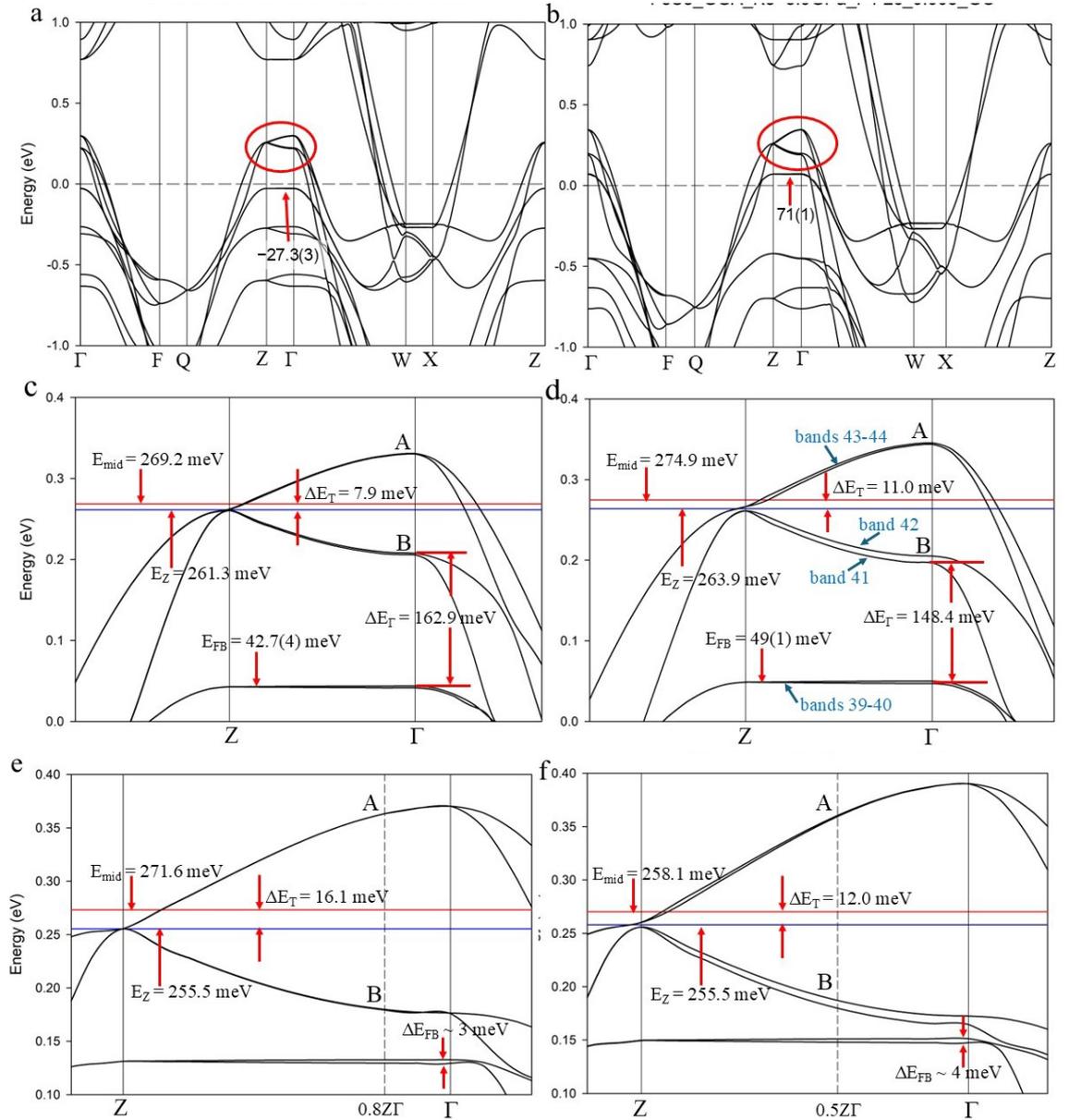

**Figure 3:** Calculated band structures for the 2*c* superlattice (space group P1) based on experimental data[3, 7] for orthorhombic β–FeSe$_{1-x}$ (a) at 5 K and 0 GPa and (b) at 8 K and 3.9 GPa using the GGA functional and geometry optimisation of atomic positions. Changes in band structure along the Z–Γ direction (circled) are the focus of this study. The energy of the FB

(arrowed) near to $E_F$ (dotted horizontal line) increases with increased pressure. (c) Detailed view of Z–Γ region at 8 K and 2.3 GPa using the GGA functional and (d) at 8 K and 3.9 GPa using the LDA functional; band numbers are discussed in Section 3.3. (e) at 8 K and 7.0 GPa using the GGA functional and (f) at 16 K and 9.0 GPa[6] using the LDA functional. The FB shifts to higher energy close to, or intersecting with, the band(s) at Γ. To accommodate band interference, $E_{mid}$ is determined by positioning "A" and "B" at intermediate points along Z–Γ. Derived parameters shown in these figures and for other pressures are listed in Table 1.

Enlarged views of the Γ–Z section calculated with the GGA functional are shown in Figure 3c and with the LDA functional in Figure 3d with applied pressures of 2.3 GPa and 3.9 GPa, respectively. Figure 3c and 3d detail an approach to evaluate energy asymmetry of a folded cosine band. For example, the energy at the mid-point between the nodal points (A) and (B) of a folded cosine band is determined and shown as $E_{mid}$ in Figure 3c and 3d. The energy at the intersection of folded bands at Z, $E_Z$, is also determined and the resulting difference, $\Delta E_T$, is a measure of band asymmetry in the (unfolded) cosine-shaped band. Figure 3 also shows another derived parameter, $\Delta E_\Gamma$, which measures the energy difference between the FB and the lower folded band at Γ.

With additional pressure, the energy of the FB also increases such that, depending on the functional used for DFT calculation, it interacts with, or affects, the folded cosine band along Z–Γ.[27, 29] Figure 3e shows the FB in close proximity to, but not crossing, the lower folded cosine band at Γ. Figure 3f shows the FB in close proximity to the lower folded cosine band for a DFT calculation of experimental data at 16 K and 9.0 GPa using the LDA functional.

**Figure S8** (Supporting Information) shows the EBS for orthorhombic β–FeSe$_{1-x}$ at 8 K and 9.0 GPa using the GGA functional. In this instance, the FB appears to intersect the lower folded cosine band. Colour coding of key bands along Z–Γ suggests that band 40 and band 41 are at almost equal energy value (within 0.6 meV) between ~0.4Z–Γ and ~0.5Z–Γ (Figure S9a). Figure S9b shows a more detailed view of the interaction of bands 40 and 41 which suggests that the bands are degenerate (to the resolution of this study). Interaction of the FB with the lower cosine band is evident and renders the methodology detailed in Figures 3e and 3f problematic for the calculation at 9.0 GPa with the GGA functional.

**Figure 4** tracks the change in energy value for FBs and for the lower folded band using calculations with LDA and GGA functionals at applied pressure. At 9.0 GPa, FBs show energy values similar to, or greater than, the energy of the lower folded cosine band. Energy convergence at ~9 GPa for structural data obtained at different temperatures (i.e. 8 K[7] and 16

K[6]) coincides with the peak for experimentally determined $T_c$ values.[8] Notwithstanding the low abundance of Cmma β–FeSe$_{1-x}$ at higher pressures (e.g. 18 GPa to 23 GPa[7]), DFT calculations show distorted cosine bands in this Z–Γ region and are discussed further in Section 3.2. Interaction of flat band(s) with the lower folded cosine curve requires adjustment of gap calculation(s) in order to estimate values for $T_c$.

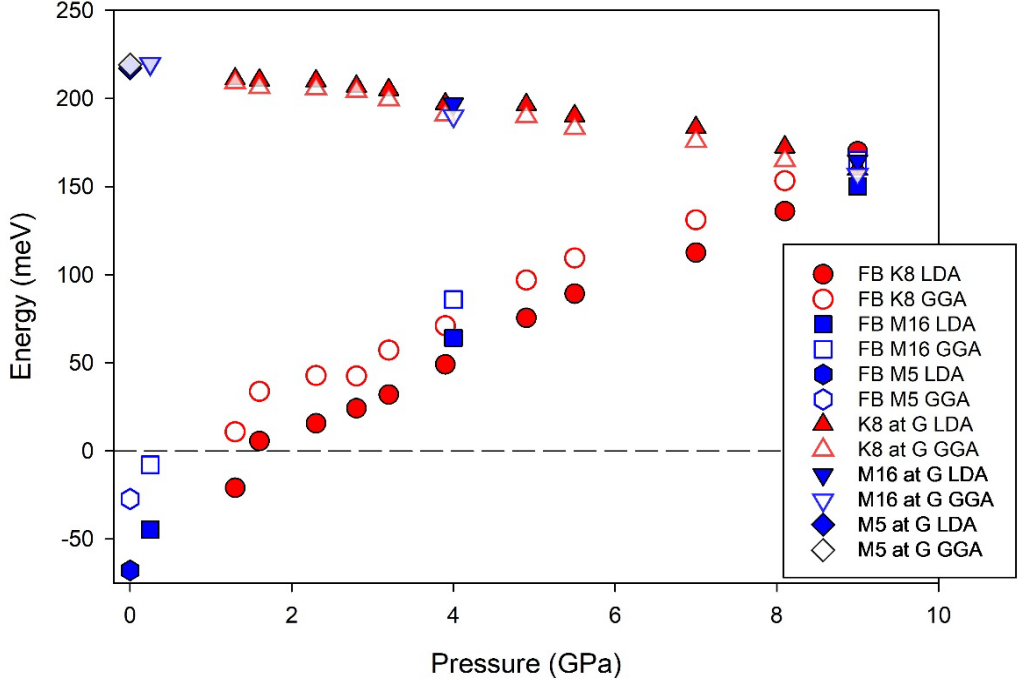

**Figure 4:** Energy values from DFT calculated EBSs for β–FeSe$_{1-x}$ (P1 2$c$ lattice) with pressure showing the lower band of the folded cosine curve intersection with the Γ node (labelled "B" in Figure 3c,d) as triangles and diamonds and the FBs as circles, squares and hexagons. Data sources are for K8,[7] M5[3] and M16.[6] Calculations for low pressures (P < 1.6) GPa show FBs along Z–Γ are below the Fermi level (horizontal dotted line). Calculations using the LDA functional are solid symbols, and open symbols are for the GGA functional.

2.2.3 Estimates for $T_c$

The $T_c$ for β–FeSe$_{1-x}$ at pressure is estimated using the value for $\Delta E_T$ obtained from DFT calculations with P1 symmetry for a 2$c$ superlattice as shown in Figures 3c,d and 3e,f. A primary reason for use of the primitive structure is the relative absence of other intersecting bands with the folded cosine curve or of bands that cross the cosine curve. Calculated $T_c$ values are based on the DFT-derived values for $\Delta E_T$ shown in Figure 3 with an adjustment for the temperature at which the crystal data are collected. For example, adjustments to $\Delta E_T$ of 0.431 meV, 0.689 meV and 1.379 meV at 5 K, 8 K and 16 K, respectively, give the values in Table 1 for $\Delta E_{T+t}$. Determination of $T_c$ includes attribution of a thermal energy to the available degrees of freedom equivalent to the energy asymmetry of the cosine-shaped band. Hence, superconductivity is implicitly associated with the thermal energy that can be absorbed and transported *via* coherent nesting which is described in Section 2.3.

As detailed in Methods, DFT calculations utilise refined cell dimensions determined from low temperature powder diffraction experiments on samples with > 50% β–FeSe$_{1-x}$.[3, 6, 7] Values for derived parameters, E$_{FB}$, ΔE$_T$, ΔE$_\Gamma$, T$_c$ and error estimates, using the calculated EBS for a P1 2c lattice are compiled for 0.0 ≤ P ≤ 9.0 GPa in **Table 1**.

**Table 1.** Starting and derived parameters for DFT calculations of orthorhombic β–FeSe$_{1-x}$.

| T$_{meas}$* (K) | P$_{meas}$* (GPa) | a = b (Å) | c (Å) | Υ (º) | Av Flat Band (meV) | Flat Band SDev+ (meV) | Av ΔE$_{T+t}$ (meV) | ΔE$_{T+t}$ SDev+ (meV) | Av T$_c$ (K) | T$_c$ SDev+ (K) | Av ΔE$_\Gamma$ (meV) |
|---|---|---|---|---|---|---|---|---|---|---|---|
| 8 | 1.30 | 3.73297 | 10.65660 | 89.664 | -5.10 | 0.80 | 7.10 | 0.71 | 17.74 | 1.77 | 218.1 |
| 8 | 1.60 | 3.71486 | 10.58820 | 90.240 | 19.65 | 0.40 | 7.82 | 0.82 | 19.54 | 2.05 | 190.9 |
| 8 | 2.30 | 3.70772 | 10.55220 | 90.114 | 29.15 | 0.40 | 7.85 | 0.91 | 19.61 | 2.26 | 178.7 |
| 8 | 2.80 | 3.70160 | 10.51300 | 90.253 | 32.90 | 0.40 | 9.10 | 1.00 | 22.79 | 2.50 | 167.3 |
| 8 | 3.20 | 3.69560 | 10.48240 | 89.721 | 44.50 | 0.40 | 9.85 | 1.06 | 24.61 | 2.65 | 157.7 |
| 8 | 3.90 | 3.68306 | 10.37540 | 89.571 | 60.00 | 1.00 | 12.61 | 1.26 | 31.51 | 3.14 | 133.9 |
| 8 | 4.90 | 3.66271 | 10.33020 | 90.006 | 86.20 | 0.45 | 12.42 | 1.39 | 31.02 | 3.48 | 107.1 |
| 8 | 5.50 | 3.65210 | 10.29460 | 89.745 | 99.35 | 0.50 | 15.04 | 1.61 | 37.58 | 4.03 | 87.3 |
| 8 | 7.00 | 3.63360 | 10.19900 | 89.974 | 121.80 | 0.45 | 15.55 | 1.63 | 38.85 | 4.06 | 57.9 |
| 8 | 8.10 | 3.61581 | 10.14040 | 89.664 | 144.65 | 0.80 | 14.98 | 1.38 | 37.41 | 3.44 | 24.2 |
| 8 | 9.00 | 3.58910 | 10.03380 | 89.738 | 178.2 | 12.1 | 12.9 | 4.3 | 32.2 | 10.8 | -22.1 |
| 5 | 0.00 | 3.76254 | 10.97200 | 89.716 | -47.55 | 0.30 | 4.18 | 0.40 | 10.43 | 1.01 | 265.7 |
| 16 | 0.25 | 3.74831 | 10.90900 | 89.877 | -26.25 | 0.35 | 4.50 | 0.42 | 11.24 | 1.06 | 246.2 |
| 16 | 4.00 | 3.67178 | 10.38860 | 89.608 | 75.00 | 0.75 | 13.57 | 1.32 | 33.90 | 3.29 | 118.3 |
| 16 | 9.00 | 3.60488 | 10.06080 | 89.664 | 157.50 | 1.50 | 14.25 | 1.20 | 35.60 | 3.00 | 7.7 |

*Values from references[3, 6, 7]; +standard deviation; derived values for E$_{FB}$, ΔE$_{T+t}$, ΔE$_\Gamma$, and T$_c$ are averages based on LDA and GGA calculations. ΔE$_{T+t}$ is the value of ΔE$_T$ with adjustment for the temperature at which crystallographic data are collected.

To convert ΔE$_{T+t}$ values to T$_c$, we use an approximation based on BCS theory[36] which, in general, shows that the gap at T = 0 K is comparable to k$_B$T$_c$[36, 37] (where k$_B$ is Boltzmann's constant) with a numerical factor of 1.764 at T(0). Experimental values of 2Δ for a wide range of materials (mostly weakly coupled) in different k-space directions generally fall between 3.0 k$_B$T$_c$ and 4.5 k$_B$T$_c$.[37] In this work, a linear extrapolation of values for 2Δ and T$_c$ gives a conversion factor of 4.65 k$_B$T$_c$ at the calculated pressures. The average values (with standard deviation) shown in Table 1 are derived from EBSs calculated with GGA and LDA functionals. At pressures where the folded cosine curve appears affected by higher energy flat band(s), a modified approach to determination of ΔE$_T$ is implemented. The determination of E$_Z$ remains as for lower pressures. The energy for E$_{mid}$ is determined at a specific point along the reciprocal Z–Γ direction depending on the pressure and temperature of data collection. For example, at 7.0 GPa the influence of the FB is considered nominal as the FB is ~58 meV lower in energy than the lower folded cosine band. In this case, "A" and "B" are positioned at a reciprocal point 20% from the Γ node towards the Z nodal point as shown in Figure 3e. For the calculation at 16 K and 9.0 GPa, the determination of $E_{mid}$ energy is at 0.5Z–Γ (Figure 3f).

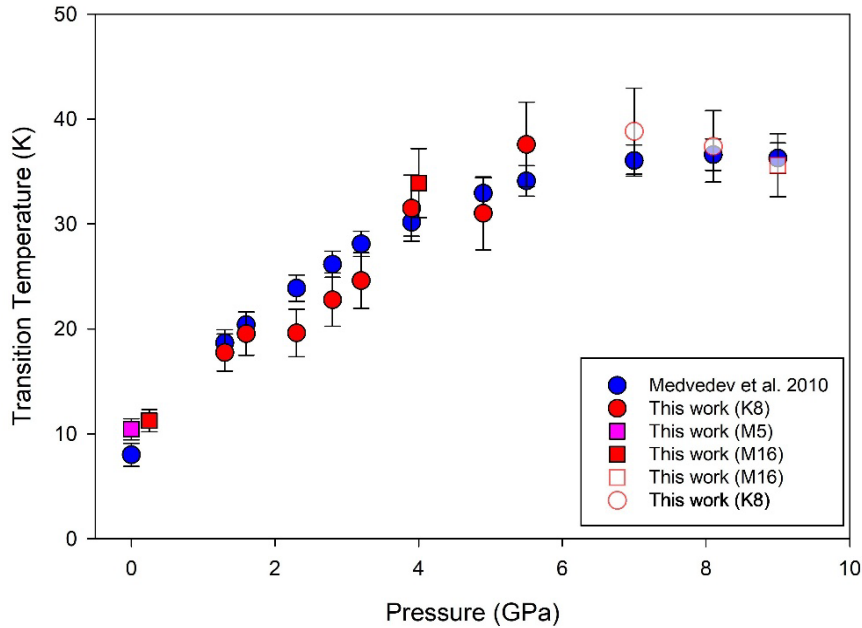

**Figure 5**: Comparison of experimental $T_c$ (blue circles),[8] with average calculated $T_c$ (red and pink symbols) using DFT with GGA and LDA functionals for the P1 2$c$ superlattice cell for β–FeSe$_{1-x}$. Parameters derived from GGA and LDA calculations are averaged to provide estimates of calculated $T_c$. Error bars are determined from these two types of DFT calculations on the same crystallographic data set. Calculated estimates for $T_c$ align well with experimental data. Calculated $T_c$ at higher pressure (i.e. ≥ 7 GPa; open symbols) are modified to account for FB interaction.

**Figure 5** compares experimental data compiled by Medvedev et al.[8] on β–FeSe$_{1-x}$ for x = 0.01 with estimated $T_c$ values determined by DFT calculations based on experimental diffraction data used in this work. In general, the match of calculated values with experimental values for pressures up to ~6.0 GPa are good, and with adjustment for FB interaction at higher pressure, calculated values are also consistent with experimental data. The estimated error for $T_c$ shown in Table 1 for the calculation at 8 K and 9.0 GPa is large and arises from substantial interaction of the FB with the cosine band using the GGA functional (see Figure S8). This specific value is not included in Figure 5. Considering the systematic errors in experiments undertaken at high pressure[38] and for DFT calculations,[32] the comparison in Figure 5 engenders confidence in DFT calculations.

## 2.3 Fermi Surfaces

In Figure 3d, bands 41-42 and 43-44 are folded representations of the EBS along the $k_z$ direction for the orthorhombic Cmma and the primitive P1 1c structures shown in Figures S6 and S7b, respectively. **Figure S9** (Supporting Information) shows calculated FSs of cosine-shaped bands 41-42 and 43-44 for the $2c$ primitive cell of β–FeSe$_{1-x}$ at various pressures. With increasing pressure, the FS and nesting vectors for bands 41-42 and 43-44 increase in amplitude as shown in Figure S9a and S9b. In addition, the truncated section of the tubular shape changes slope with change in pressure. Figure S9c shows Fermi surfaces of bands 43-44 with nesting vectors that show large sections of the Fermi surface are connected by the same vector.

Figures S9d and S9e illustrate the influence of an FB at the Fermi surface in relation to that for bands 41-42. As applied pressure increases, the FBs (bands 39-40) increase energy and for P < 5.5 GPa, the FS is unremarkable. Above 5.5 GPa, the FSs for bands 39-40 not only increase in amplitude but also transform to a convex tubular shape similar to that for bands 41-42 at lower pressure (P < 5.5 GPa) and bands 43-44 at higher pressure (P > 7.0 GPa). At 9.0 GPa and up to 21 GPa, this convex tubular shape is more pronounced for bands 39-40. The Fermi surface of the FB collapses at 23 GPa (Figure S9e).

### 2.3.1 Isoenergetic Fermi Surfaces

**Figure S10** (Supporting Information) shows the sensitivity of FSs to small variations in potential for β–FeSe$_{1-x}$. In general, the FSs for all bands near $E_F$ (i.e. bands 39-40, 41-42 and 43-44) show variations to topology with increments of 50 meV that are similar to topological changes with pressure (Figure S9). For example, bands 39-40 at 1.6 GPa show similar topology to bands 39-40 with application of 100 meV potential; yet with −50 meV potential, the FSs of bands 39-40 are similar to an applied pressure of 8.1 GPa at 8 K. These changes to FSs near $E_F$ are discussed in Section 3.4

### 2.3.2 Bands parallel to Γ–Z offset along $k_y$

Typically, electronic bands are calculated for primary crystallographic orientations as shown in Figure 3. Calculations on bands parallel to a primary axis, at samplings offset along a direction parallel to c$^*$ and along $k_y$, for example, also reveal electron behaviour close to $E_F$.[39] **Figure 6** shows bands for values of $k_y$ along Γ–A for calculated band structures parallel to Γ–Z. These calculations for β–FeSe$_{1-x}$ at 2.8 GPa and 9.0 GPa, as well as for examples in **Figure S11** (Supporting Information), are undertaken using the LDA functional with input parameters detailed in Methods.

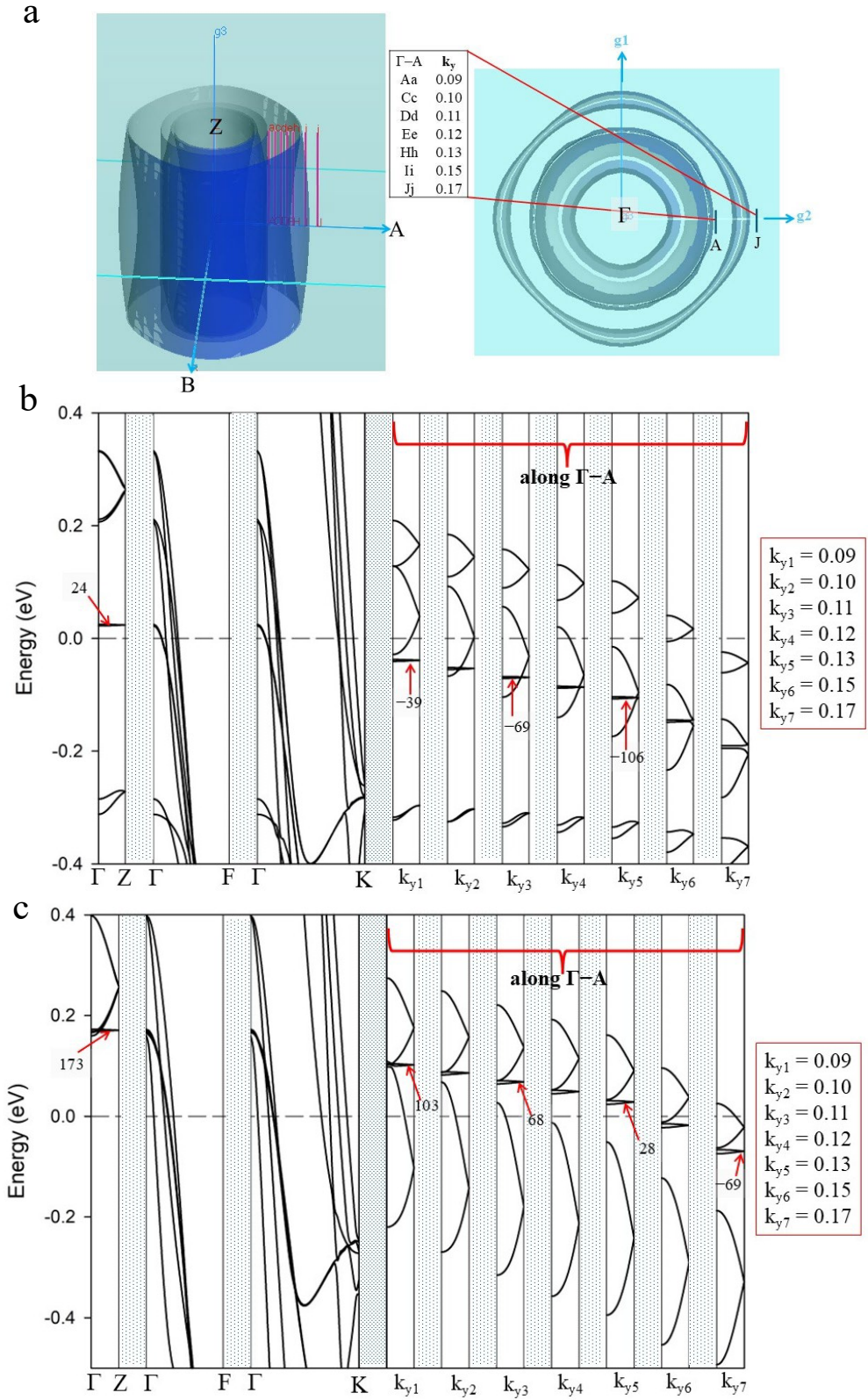

**Figure 6.** Band features near $E_F$ for orthorhombic β–FeSe$_{1-x}$ (a) Fermi surfaces showing reciprocal orientations and $k_y$ sampling locations along Γ–A. (b) EBS calculated using the LDA functional for a P1 2*c* lattice at 8K and 2.8 GPa and (c) at 8 K and 9.0 GPa. Left hand panels (Γ to K) show high symmetry directions as presented in Figure 3. Right hand panels (designated

$k_{y1}$, $k_{y2}$ etc) show band formats at specific locations offset along Γ–A (i.e. parallel to Γ–Z between 0, $k_y$, 0 and 0, $k_y$, 0.5). Energy values for the FBs are in meV. For P>7.0 GPa, FB splitting at Γ ranges from 3 meV to ~8 meV so, energy values for $\mathbf{k_y}$ are at the intersection with the Brillouin zone boundary plane containing Z.

The EBSs in Figure 6b and 6c show bands for conventional directions (as per Figure 3) and, in the right hand panels (designated $k_{y1}$, $k_{y2}$ etc), bands at specific points within the first Brillouin zone offset along Γ–A. At 2.8 GPa and 9.0 GPa, bands 43-44 and 41-42 along Γ–Z are degenerate as shown on the left hand panels of Figure 6b and 6c. Offset along Γ–A, the formerly degenerate bands split into two bands above and below $E_F$. A third band with a narrow energy range, which represents lone pair electrons, is also present at these sampling locations in reciprocal space. At 2.8 GPa for $k_y = 0.13$, the split bands shift to separate bands above and below $E_F$ while at 9.0 GPa the separation occurs at $k_y = 0.12$.

The lone pair FBs cross $E_F$ along Γ–A at different values for $k_y$ (0.10 and 0.15 for 2.8 GPa and 9.0 GPa, respectively) and are at energy values that intersect, or are similar to, one of the split bands. As anticpated, the energies of FBs at 9.0 GPa are higher than that at 2.8 GPa. At 9.0 GPa, the FB appears pinned to the lower section of the split conduction band not only at the Γ node for Γ–Z but also at parallel reciprocal directions offset along Γ–A. In Figure 6c, the FB at all samplings offset along Γ–A shows a detectable split of energy (nearest to Γ) ranging from 3.8 meV to 8.0 meV between $k_{y1}$ and $k_{y7}$. Non-degenerate FBs at the plane perpendicular to the ΓZ axis containing Γ are of similar order with increased applied pressure (Figure S11b), albeit reverting to degeneate energy band(s) at the Brillouin zone boundary plane containing Z. Transition of FBs from degenerate to non-degenerate along along Γ–Z is less evident in EBSs at lower pressure. Along Γ–Z, the FB is degenerate (within < 1 meV) at lower pressure (i.e. ≤ 4 GPa) and, on average, shifts from below $E_F$ to above $E_F$ at ~1.6 GPa.

2.4 Lattice Vibrations

Calculations of lattice vibrations for the tetragonal and orthorhombic structures of β–FeSe$_{1-x}$ show similar vibration modalities with applied pressure. Two vibration modes invoke movement of interlayer Se atoms in opposite directions while all other modes are articulated movements of intralayer atom groups or their combinations in the same direction(s). Interlayer Se modalities for tetragonal and orthorhombic structures, at 300 K and 8 K respectively, are shown in Supporting Information (**Figure S12**). Acoustic frequencies are due to opposing Se movements within the *x-y* plane and optical frequencies refer to opposing Se movement along the $c$ axis. These calculated vibration trends for tetragonal and orthorhombic structures are consistent with phonon density of states experiments that track spectral behaviour with change in temperature across the structural transformation[17, 40] and with the broad consensus that electron pairing is not driven by lattice phonons in FeSCs.[26]

3. Discussion

In this work, the focus is on utilising experimental data collected with the same equipment across the range of pressures and temperatures (i.e. 1.3 GPa < P < 23 GPa; 5 K ≤ T ≤ 16 K) for which orthorhombic β–FeSe$_{1-x}$ (SG: Cmma) occurs.[7] The percentage of Cmma phase, initially 97% at 1.3 GPa and 8 K,[7] reduces substantially at pressures > 12 GPa[7, 8] with complete transformation to the Pbnm phase above 26 GPa.[7] At 15 GPa, the T$_c$ for β–FeSe$_{1-x}$ (x = 0.01) drops to ~25 K but with a resistivity measure that is not zero due to the increased presence of a second phase.[8] Experiments at higher pressures[8] show that superconductivity persists with reduced T$_c$ values up to ~ 20 GPa as indicated in Figure 1. Crystallographic experiments also show that at 26 GPa the proportion of Cmma phase reduces to ~7.6% and that of the Pbnm phase increases to 92.4%.[7] Experimental data show that the phases above ~20 GPa (hexagonal or orthorhombic Pbnm) do not superconduct.[7, 8]

3.1 Bonding

In this section, changes to the orthorhombic Cmma structure with applied pressure and the effects on bonding are briefly explored. Covalent bonding between Fe and Fe, as well as Fe and Se, for tetragonal and orthorhombic forms of β–FeSe$_{1-x}$ is established.[6, 8] For Fe–Se bonds, the difference in electronegativity values for each atom (δX = 0.72) suggests that the hybrid sp$^3$ Fe–Se bonding is polar with a higher partial charge on Se than on Fe.[41, 42] In octahedral β–FeSe$_{1-x}$, each Se has four surrounding intralayer Fe atoms. Recent estimates for electronegativity in solids that take into account the number of bonding electrons and coordination number,[43] gives a δX value of 0.93. Both values for δX in octahedral β–FeSe$_{1-x}$ indicate polar covalent Fe–Se bonds.

3.1.1 Electron density

The structure of a solid crystal can be substantially influenced by the presence (or absence) of lone pairs.[44] Selenium displays non-bonding character in crystals, leading to the presence of lone pairs of electrons, important for many chemical, surface and electronic properties.[44-47] The VSEPR theory[48] posits that all electron groups—both bonding pairs and lone pairs—will repel each other and arrange themselves to be as far apart as possible. In general, lone pairs exert a greater repulsive force than bonding pairs[49] and can influence covalent bond polarity. Selenium has a $4s^24p^4$ configuration ([Ar core] $4s^23d^{10}4p^4$) with an electron in two p orbitals to share covalent bonds with Fe in β–FeSe$_{1-x}$. With six valence electrons, other orbital electrons form two lone pairs per atom. This type of bonding and non-bonding configuration generally results in a "bent tetrahedral" geometry around the Se atom. The degree of bend depends on the repulsive force of lone pairs on the Se nucleus.

A symmetric tetrahedral geometry (i.e. without a lone pair) has a tetrahedral angle of 109.5°. This value compares to a bent tetrahedron that ranges from 104.5° to 103.3° in tetragonal β–FeSe$_{1-x}$ at 300 K with pressure and, for orthorhombic β–FeSe$_{1-x}$ at 8 K, from 106.2° to 101.0° for 1.3 GPa < P < 9.0 GPa. In both structures, this bent angle circumscribes an interlayer distance. The changes in tetrahedral angles with pressure for tetragonal and orthorhombic structures are compared in Figure S4d (Supporting Information). For the orthorhombic structure, the tetrahedral angle changes by ~5% for 1.3 GPa < P < 9.0 GPa compared to a ~1% change in the tetragonal structure for 0 GPa < P < 11.0 GPa. The relative distortion of tetrahedral geometry under conditions of low temperature (≤ 16 K) and pressure, facilitated by lone pair repulsions, is an important influence on electronic properties of orthorhombic β–FeSe$_{1-x}$.

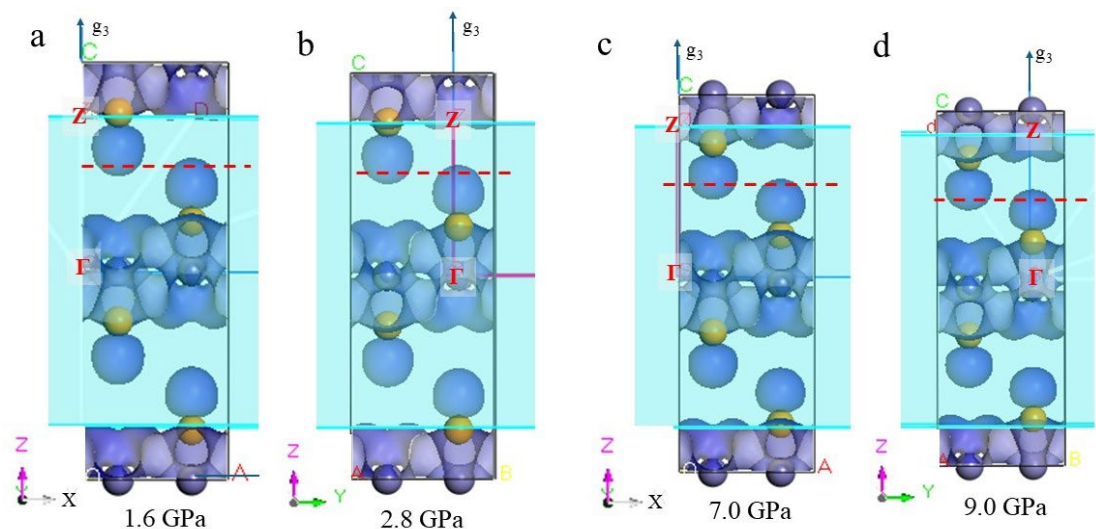

**Figure 7:** EDD plots for orthorhombic β–FeSe$_{1-x}$ calculated for a 2*c* P1 lattice at (a) 1.6 GPa, (b) 2.8 GPa, (c) 7.0 GPa and (d) 9.0 GPa showing projections down the *b* axis. Fe–Fe and Fe–Se bonds are shown by blue and dark blue shading. Se atoms (yellow) also show localised electron density, or lone pair electrons, in the interlayer region. Note that the lone pair regions move closer to the mid-point of the Se planes with increased pressure. The dotted red line is at the mid-plane between Se atoms as an aid to the eye.

**Figure 7** shows the electron density difference[50] (EDD) for orthorhombic β–FeSe$_{1-x}$ at a range of pressures calculated using the LDA functional. Each EDD shows the interlayer region parallel with the *c* axis. Figure 7 reveals high EDD directly associated with the interlayer Se atoms. We consider these electron dense regions represent projections of lone pair electrons, often called non-bonding electrons. With increased pressure, these lone pair electrons shift closer to the mid-plane between Se atoms and are a significant driver of structural adjustments with pressure. At 9.0 GPa, these lone pair electrons extend past the mid-plane clearly demonstrating a change in electron distribution along the (real space) *c* axis direction with increased pressure.

3.1.2 Orbital overlap

In crystalline solids, atomic orbital interactions determine the conditions for bonding in crystals. These interactions include the degree and extent of orbital overlap as well as similarity of orbital energies and correct symmetry.[51] Molecular orbital calculations also enable computational methods to extend density of states (DOS) and partial density of states (PDOS) concepts through the use of crystal orbital overlap population (COOP) analysis.[52] This chemical bond analysis is based on an extension of the Mulliken population analysis using molecular orbital calculations. In COOP analysis, orbital overlap is quantified in order to delineate chemical bonding based on calculations with atom-centered orbitals.[52]

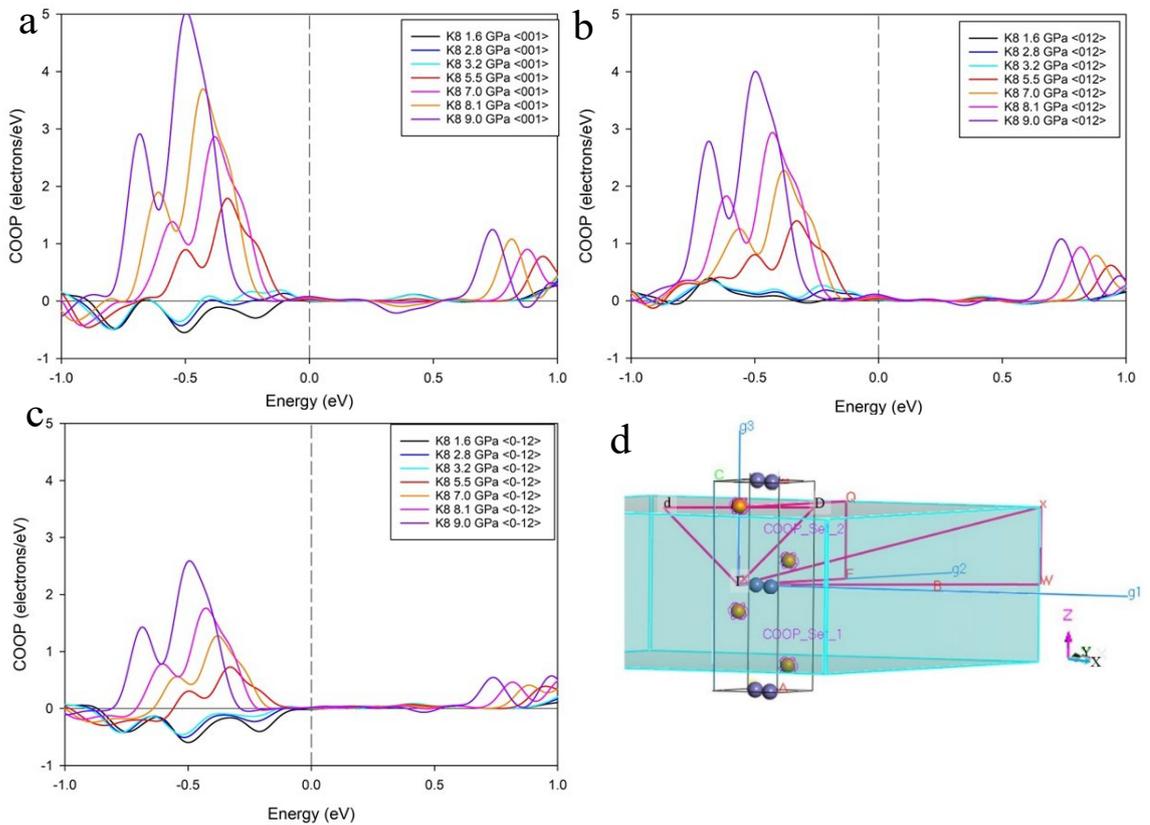

**Figure 8:** COOP plots for orthorhombic β–FeSe$_{1-x}$ with a P1 2$c$ lattice calculated using DMol3 for 1.6 GPa < P < 9.0 GPa (a) for Se–Se atoms parallel to <001>; (b) Se–Se atoms parallel to <012>; (c) Se–Se atoms parallel to <0$\bar{1}$2> and (d) schematic showing real space orientation of atom sets used to calculate COOP values for β–FeSe$_{1-x}$.

**Figure 8** shows COOP analyses of atom–atom bond sets for β–FeSe$_{1-x}$ with directions chosen parallel to <001>, <012> and <0$\bar{1}$2> to include the interlayer area of the unit cell shown in Figures 2 and S4. Figure 8d shows the real space orientation for the COOP projections in the P1 2$c$ primitive unit cell. The projections in Figure 8 indicate that interlayer Se–Se bonding shifts from antibonding (negative values) to bonding near E$_F$ at lower pressure (e.g. < 4.0 GPa). For P > 4.0 GPa, bonding orbitals (positive values) predominate close to E$_F$ with a shift to lower energy values as pressure is increased. Furthermore, the relative proportions of bonding and antibonding orbital contributions changes with pressure.

3.1.3 van der Waals Bonds

VdW forces result from transient shifts in electron density and are important to consider for polar bonds, highly polarizing atoms and/or the presence of lone pairs where electron density may shift proportionately around a nucleus. Calculations for vdW radii in crystalline forms of Se and Fe, where there is neither a repulsive nor attractive force, estimate 1.9 Å and 2.05 Å, respectively.[53] The presence of lone pairs on an atom can contribute to, or affect, the formation of vdW bonds,[44] as well as influence polar covalent bonds.

A Se–Se distance of 3.8 Å is a reasonable estimate for equilibrium vdW distances in solid Se compounds.[44, 53] Table S2 shows that the shortest Se–Se distances in orthorhombic β–FeSe$_{1-x}$ are interlayer distances and less than 3.8 Å with application of pressure. The interlayer Se–Se distances for both tetrahedral and octahedral structures are shorter than the equilibrium vdW value, suggesting attractive interactions that occur between the layers are more than typical vdW forces.

A detailed example of this effect has been postulated for layer structured Sb$_2$Te$_{3-x}$Se$_x$. This study notes that the underlying bonding may be partly responsible for key properties that are not observed for materials with classic vdW bonds.[46] For Sb$_2$Te$_{3-x}$Se$_x$, homo-atomic vdW-like bonds (e.g. Se–Se) mainly contribute to the $c$-axis lattice parameter; similar in character to the interlayer Se–Se distances in Table S1 and Table S2 for β–FeSe$_{1-x}$. However, Figure 8 shows that interlayer Se–Se bonding occurs near E$_F$ and at energies for E$_F$ > 0. We suggest that vdW bonding in the β–FeSe$_{1-x}$ structure does not collapse,[8] but adjusts to structural rearrangements, facilitated by the presence of non-bonding, or lone pair, orbitals in the transition to an orthorhombic structure. These adjustments to bonding are further moderated by increasing polarity between Fe and Se atoms with pressure.

3.2 Fermi Surfaces

The concave tubular shaped FSs shown in Figure S9 are similar in format to that described for the layer structured superconductor, $MgB_2$[27, 39] albeit with different response(s) to increased pressure. For $MgB_2$, changes in pressure result in a narrowing of the convex tubular shape FS at Γ and retention of a clear distinction between the light and heavy effective mass σ bands.[39] Accordingly, the $T_c$ for $MgB_2$ reduces consistently with application of pressure.[25]

For β–$FeSe_{1-x}$, bands 43-44 show varying degrees of concave inner tubular shape with an outer convex curvature for pressures up to 21 GPa. Bands 41-42 are clearly affected by the FBs (bands 39-40) at 8.1 GPa and 9.0 GPa in (Figure S8b) in that the concave tubular shape does not occur. At other pressures, similar concave inner tubular shapes and outer convex curvature shapes occur. Drawing an analogy with similar studies on $MgB_2$,[27, 32, 39] these concave and convex tubular shapes reflect the light and heavy effective mass σ bands, respectively. The sequence of Fermi surfaces with change in pressure shown in Figure S9 implies that electron transfer *via* inter-band and intra-band hopping[39] may be important mechanisms for superconductivity in β–$FeSe_{1-x}$.

Isoenergetic surfaces are closely linked to the dispersion of electronic bands[54] and play a significant role in understanding electron dynamics. Experimental determination of superconductivity requires exposure of a sample to an external electromagnetic field resulting in a range of isoenergetic FSs.[54] Distinct changes to FSs for bands near $E_F$ as shown in Figure S10 suggest that, for example, a −50meV potential in the near vicinity of the FB (at 7.0 GPa) incurs increased definition and expansion of tubular shaped surfaces. In this case, the isoenergetic surfaces suggest that localised electrons may gain (or lose) charge under specific geometric conditions. At high pressure, this delocalisation implies conjugate π bonds or a probable resonance stabilised bonding in the interplanar regions of β–$FeSe_{1-x}$. In contrast, a similar applied negative potential to the (folded) cosine bands (bands 41-42 and 43-44) shows minimal effect whereas a +50meV potential enhances the tubular shapes reflecting light and heavy effective mass σ bands.

3.3 Flat Bands

A FB is generally formed from states that have limited, if any, orbital overlap between atoms on lattice sites. Flat bands are characterized by an energy dispersion independent of momentum and with zero kinetic energy of electrons.[55]. At the atomic limit, a flat dispersion in momentum *k* space suggests that an electron cannot hop between atom sites in a crystal.[56] More recent interest in flat, or zero dispersion, bands arises from observations that hopping between sites can occur in particular lattice configurations.[56] Detailed analyses of bulk crystalline structure databases have identified a wide range of FB lattices in 3D materials,[57-59] particularly those with two or more sites in a unit cell.[55]

For crystalline materials, database analyses[57, 58, 60] have identified lattice and orbital characteristics that indicate potential for non-trivial FBs near the Fermi surface in either all, or part of, the Brillouin zone. For example, common FB lattices comprise *p* and *d* block elements[58] and if a sublattice is weakly perturbed or deformed by remaining atoms or orbitals, Flat Topological Bands (FTBs) may be present.[57] FTBs occur when Bloch waves are extended in one (or more) lattice direction(s) resulting in quenched kinetic energy due to interference effects despite electron orbital overlaps and hopping.[57] In the limit where the kinetic energy of electrons is quenched, modest electron-electron interactions can dominate with resultant unconventional ground states.[55]

3.3.1 Flat bands with Pressure

Calculations based on experimental data show that FBs in orthorhombic β–$FeSe_{1-x}$ respond to applied pressure. With increased pressure, the calculated FB energy increases to values similar to that for the lower cosine band at the Γ nodal point of the 2c lattice. As shown in Figure 3d, when the FB energy is within ~60 meV of the value at the Γ nodal point, localised electrons appear to influence bonding electrons.

We suggest that FBs in calculated band structures for β–$FeSe_{1-x}$ are due to the presence of lone pair, or non-bonding, orbitals consistent with earlier studies on chalcogens.[47, 61, 62] Calculations with increased pressure show that the lone pair electrons localised to Se atoms exert geometric constraints on bonding – that is, on the electron distribution – within the orthorhombic structure. The EDD projection along the $c^*$ axis shown in Figure 8 exemplifies this condition where the relative positions of localised electrons within the interlayer space change with pressure. Consistent with this, a substantial shift in *c* axis dimensions (e.g. Se–Se distances; Table S2) and reduced tetrahedral angle (e.g. Figure S4d) occurs with increased pressure. The effect of FB proximity is also three dimensional; evident by the orthogonal projections along Γ–A in Figure 6.

3.3.2 Flat bands and Superconductivity

The peak value for experimentally determined $T_c$ is 36.7 K at 8.9 GPa;[8] a pressure at which the DFT calculated energy of FBs (using either LDA or GGA functionals) matches, or interferes with, the lower branch of the folded band at the Γ nodal point of a $2c$ superlattice. Calculations at higher applied pressure also reveal the influence of lone pair electrons in β–FeSe$_{1-x}$. For example, FB interference with cosine shaped bands is evident when calculated for 9.0 GPa using the GGA functional as shown in Figure S8. In this case, band trajectories suggest that FBs 39-40, degenerate at and near nodal point Z, lose degeneracy closer to Γ along along lines parallel to Z–Γ. Close examination (Figure S8b) shows that FB 40 intersects with lower cosine band 41 (becoming degenerate for a brief segment of reciprocal space) and, as with band 39, shifts to dispersive band(s) near the Γ nodal point. Concommittently, bands 41-42 show a flat non-degenerate band character from the intersection of bands 40 and 41 to the Γ nodal point.

At higher pressures (P > 9.0 GPa; **Figure S13**, Supporting Information), partial FBs along Z–Γ not only disrupt the dispersive cosine bands that occur at lower pressures but transform to separate non-contacting, non-crossing bands. At these higher pressures, FBs split into separate degenerate and partially degenerate partial FBs nearest to the Z and Γ nodal points, respectively. Consistent with calculations at lower pressure (i.e. < 9 GPa), the average energy of these partial FBs also increases with applied pressure.

One possible interpretation of the above observations is that a different superconducting mechanism occurs for orthorhombic (Cmma) β–FeSe$_{1-x}$ at high pressures (e.g. ≥ 15 GPa) albeit with a lower T$_c$. However, this inference need not be correct. For example, at higher pressures, charge transfer is even further moderated by geometric factors due to, for example, closer interaction of lone pair electrons. Charge transfer, at pressures insufficient to force complete structural transformation to the Pbnm phase, may be further induced by a degree of electron de-localisation in the interlayer region of β–FeSe$_{1-x}$.

Observations that support pressure-induced electron de-localisation aided by lone pair electrons include: (i) the change in band structure along Z–Γ (compare Figures 3e, 3f and S8 with Figure S14), as well as bands offset along Γ–A near E$_F$; (ii) the shift of Se-centric EDD closer to the $c$-axis mid-point with increased pressure (Figure 7) and (iii) anti-bonding and bonding orbitals in close proximity to E$_F$ (Figure 8). The interaction of lone pair electrons with applied pressure clearly influences superconducting behaviour for P < 9.0 GPa and to a greater degree for P > 15 GPa.

3.4 Hopping Mechanisms

The electronic transport properties of solids are determined by a narrow window of energies, of order the thermal energy, at the Fermi level.[63, 64] Electron energies involved in transport phenomena belong to the window of energy at the step transition of the Fermi-Dirac distribution (between fully populated and fully unpopulated electronic states below and above the Fermi level, respectively). Cosine shaped bands of orthorhombic (Cmma) β–FeSe$_{1-x}$ display bonding/antibonding energy asymmetry with strong correlation to T$_c$ as shown above. This characteristic has been linked to a hopping transport mechanism that has potential generic application for the interpretation of superconducting band topology.[39]

Other orientations in close proximity to the Fermi surface, besides high symmetry orientations along principal real space directions (e.g. Figure 3), may also provide useful detail on electronic properties.[39] For example, at 2.8 GPa the FB is ~183 meV lower in energy than the degenerate cosine band along Γ–Z (Figure 4 and Figure 6). At intermediate locations parallel to the c$^*$ axis but offset from Γ within the Brillouin zone, the FB is close to the lower split band (within 10 meV) at k$_y$ = 0.09 below E$_F$. The FB intersects the lower split band below E$_F$ for k$_y$ > 0.09 (Figure 6). The topology of bands in the directions offset along Γ–A implies that the FB affects bonding orbitals and bonding properties in β–FeSe$_{1-x}$.

At 9.0 GPa, the FB intersects the lower part of the degenerate band along directions parallel to Γ–Z in the conduction layer. When the degenerate band splits to above and below E$_F$, the FB energy is similar to (i.e. < 3 meV) the bottom point (at Γ) of the upper split cosine band at intermediate locations orthogonal to the c$^*$ axis within the Brillouin zone (Figure 6c). At intermediate pressures (e.g. 5.5 and 7.0 GPa; Figure S11), trends for cosine band and FB topologies are similar.

This topology between the FB and the cosine band along Γ–Z suggests that charge transfer mechanisms are enhanced in β–FeSe$_{1-x}$ by the presence, and geometric confinement, of FBs with applied pressure. Thus, the band and FS topology of β–FeSe$_{1-x}$ show that a quasiparticle hopping mechanism for charge transfer (partly influenced by FBs), consistent with similar analyses on binary solids, is evident.[39]

3.5 Chalcogenide superconductors

For β–FeSe$_{1-x}$ low temperature, ambient pressure experiments using Bogoliubov quasiparticle interference (BQPI) imaging revealed the existence of orbital-selective Cooper pairing based preferentially on the d$_{yz}$ orbitals of iron atoms.[65] Modified tight binding equations to define orbital priorities and quasi-particle correlations, enabled modelling of charge transfer (or quasi-particle hopping) to match experimental data from ARPES and BQPI.[65] Orbital-selective Cooper pairs predominantly associated with the Fe d$_{yz}$ orbital in orthorhombic β–FeSe$_{1-x}$ have been further explored by Kreisel et al.[66]

These investigations focus on intralayer Fe–Fe interactions and their orbitals in orthorhombic β–FeSe$_{1-x}$ and recognise three key bands (for k$_z$ = 0) α, ε and δ[26, 65, 66] with three corresponding zero energy Fermi surfaces with a k$_z$/π component.[66] Two gaps were detected by experiment with maximum values of 2.3 meV and 1.5 meV for the α and ε bands, respectively. A gap value for the δ band could not be determined using BQPI analysis.[65] Nevertheless, a total gap of ~3.8 meV is consistent with T$_c$ = 8.7 K at ambient pressure using experimental techniques of high precision. The key role of Fe d$_{yz}$ orbitals suggests a substantial $c$ axis component that includes Fe–Se bonds with hybrid orbitals. Using unit cell parameters from Margadonna et al.[3] at 5 K and ambient pressure, the average gap value using cosine band asymmetry along Z–Γ determined with DFT calculations is 4.18 ± 0.40 meV (see Table 1).

Band structures shown in Figure 3 and Supporting Information (Figures S6–S8) focus on the c$^*$ direction of layer-structured β–FeSe$_{1-x}$. Because β–FeSe$_{1-x}$ is a stable compound at low and ambient temperatures for pressures up to > 9 GPa, interplanar bonding is key to understanding quasi-particle behaviour; despite the dimension of the $c$-axis at twice the average covalent Fe–Fe bond length.[41, 67] The Z–Γ cosine band includes quasi-particle interaction associated with orbitals for Fe–Fe and Fe–Se bonding. The latter bonding predominantly occupies the interlayer region of β–FeSe$_{1-x}$ along with lone pair electrons. The presence of lone pair electrons in β–FeSe$_{1-x}$ clearly influences behaviour of other electrons and/or quasi-particles. The approach described above has potential for enhanced understanding of superconducting mechanism(s) in the wider chalcogenide family of compounds,[68-70] as well as studies that emphasise quantum interference, quantum geometry and non-equilibrium physics of solid materials.[55, 56, 71]

4. Conclusions

*Ab initio* DFT models of orthorhombic β–FeSe$_{1-x}$ at applied pressure show that crystal structure is a dominant influence on atom-atom bonding and antibonding character evident in band structures projected along the c$^*$ direction and near E$_F$. Calculations based on experimentally determined cell dimensions at low temperature and applied pressure show cosine band asymmetry across the layer direction. Additionally, non-bonding orbitals containing localised electron pairs respond to pressure-induced structural changes that affect quasi-particle behaviour. Non-bonding, or lone pair, electrons are characterised by a flat band near the Fermi level along the c$^*$ direction in band structure plots. The energy of the flat band along Γ–Z increases with applied pressure.

Cosine bands along $c^*$, associated with tight binding models, reflect topology and energy level adjustments in response to applied pressure that causes realignments of bonding, non-bonding and anti-bonding orbitals. Cosine band energy asymmetry (equivalent to the thermal energy for the available degrees of freedom), averaged from calculations using LDA and GGA functionals, provides an excellent match to experiment with applied pressure up to ~12 GPa. Superconducting $T_c$ values for orthorhombic β–$FeSe_{1-x}$ are calculated using a primitive lattice based on Cmma symmetry with adjustment for the temperature of unit cell data collection. Similar analyses of cosine bands to determine the superconducting energy gap based on experimental unit cell data at 5 K and 0 GPa closely matches direct experimental probe(s) of β–$FeSe_{1-x}$ using Bogoliubov quasiparticle interference imaging at < 1 K.

Along Γ–Z, lone pair electrons influence bonding orbitals with increased applied pressure. At ~9.0 GPa and the maximum $T_c$ for orthorhombic β–$FeSe_{1-x}$, the flat band near $E_F$ intersects degenerate cosine bands. Analysis of calculated Fermi surfaces, electron difference distributions and bands along Γ–A parallel to Γ–Z (within the first Brillouin zone) suggests that charge transfer mechanisms are enhanced by the presence, and geometric confinement, of flat bands in β–$FeSe_{1-x}$. At pressures ≥ 18 GPa, where experimental data show orthorhombic β–$FeSe_{1-x}$ superconducts, flat bands disrupt dispersive cosine bands and transform to separate non-contacting, non-crossing, non-degenerate bands.

This approach utilising meV scale detail of DFT calculated band structures at the Fermi level extends understanding of charge transfer mechanism(s) to unconventional as well as conventional superconductors. DFT calculations on other chalcogenide superconductors using similar meV scale interpretation of band structures may provide further evidence for the impact of lone pair electrons on quasiparticle behaviour.

5. Methods

In general, computational approaches and limitations are provided in earlier publications.[27-29, 32, 35] Key input parameters for calculations are detailed below.

5.1 DFT Models

DFT calculations on the EBS and FS for β–$FeSe_{1-x}$ at pressures from 0 GPa to 23 GPa were undertaken with Materials Studio CASTEP Versions 2023 and 2025.[72] Key parameters such as plane wave cut-off energies, pseudopotentials, and k-point grids are critical enablers of meV resolution for EBS calculations.[32, 35]

Calculations of crystal orbital overlap population (COOP) utilize Materials Studio DMol3 Version 2025[73, 74] and the LDA functional in energy mode with geometry optimized input cell parameters for β–$FeSe_{1-x}$ derived from CASTEP. Calculations are with all-electron non-relativistic treatment, spin unrestricted and a global cut-off of 5.8 Å, since various tests of relativistic treatment produced identical results. Because Dmol3 cannot optimize geometry with external pressure, the recommended procedure is to optimize unit cell geometry in CASTEP and to calculate in the energy mode with Dmol3.

5.1.1 Model Interpretation

All band structure representations utilize .xlsx files converted from direct .csv outputs of Materials Studio DFT calculations. These data are then re-plotted using SigmaPlot for Windows Version 15 (Grafiti LLC, Palo Alto, CA USA) to enable precise interpretation of band energies and intersections at reciprocal lattice symmetry points. Band intersections with high symmetry points as well as degeneracies and/or band avoidances are determined by visual inspection of SigmaPlot graphs correlated with energy values to the fifth significant figure. Use of a primitive lattice, based on a known space group determination, enables precise comparison of systematic crystallographic data (e.g. changes with applied pressure) and band topology without constraints imposed by higher order symmetry conditions. This configuration is particularly important for bands close to, or crossing, the Fermi level.

Crystal structures are re-analyzed and visualized with Crystal Maker V 11.6 software (CrystalMaker Software Limited, Begbroke, Oxfordshire, UK) using cell parameters of experimentally determined values as described below. Atom-atom distances and angles are re-calculated using Crystal Maker software. Modelled crystal vibrations with force field equations use Buckingham and Lennard-Jones atom potentials for solid state inorganic materials with Monte Carlo simulation using Crystal Maker V11.6 software.[75, 76]

5.1.2 Calculation Parameters

All DFT calculations employ cut-off energies of 990 eV, $\Delta k$ grid of 0.005 Å$^{-1}$, norm-conserving pseudopotentials and the Perdew-Zunger Local Density Approximation (LDA) or Generalised Gradient Approximation (GGA) for the exchange-correlation functionals.[77, 78] Optimisation calculations used the Broyden-Fletcher-Goldfarb-Shannon (BFGS) algorithm employing a total energy/atom convergence tolerance of 0.5 x 10$^{-5}$ eV, an eigen-energy convergence tolerance of 0.7547 x 10$^{-7}$ eV, a Fermi energy convergence tolerance of 0.2721 x 10$^{-13}$ eV, a Gaussian smearing scheme of width 0.1 eV and the Pulay density mixing scheme with charge density mixing $g$ vector of 1.5 Å$^{-1}$. For optimization calculations, all atoms are allowed to relax along orthogonal directions until residual forces are less than 0.01 eV/Å. Maximum shift from special positions for P1 space group calculations is < 0.0001 Å. For all calculations, including geometry optimisation, the cell dimensions determined by structural analysis of diffraction data have been used as input parameters to DFT calculations; only site positions are geometry optimised.

## 5.2 Source Data

Detailed crystallographic studies on β–FeSe$_{1-x}$ at low temperature with applied pressure and determination of T$_c$ under the same conditions are limited due to difficulty in acquiring all three datasets simultaneously within the same experimental apparatus.[38] In some cases, studies on β–FeSe$_{1-x}$ with applied pressure are accompanied by measurement of T$_c$ on the same sample thus, directly determining temperature dependence [2, 8, 15, 40] with limited structural data and on an inferred orthorhombic form. Alternatively, crystallographic studies on β–FeSe$_{1-x}$ at specified temperature(s) with applied pressure provide an estimate of stoichiometry determined by complementary methods[4, 6, 7, 9] with inferred or limited estimates for T$_c$.

Input unit cell values are taken from experimental diffraction data for tetragonal and orthorhombic β–FeSe$_{1-x}$ powder samples.[3, 6, 7] Data are selected for this study on the basis of (i) stoichiometry with x ≤ 0.03, (ii) structural data for all detectable phases in the powder assemblage, (iii) data obtained from the same material at ambient temperature with pressure and at low temperature with pressure,[7] (iv) relative proportions of detectable phases in the powder assemblage and (v) statistics indicating the quality of crystallographic parameters at each pressure determination.

The terminology for these early datasets[3, 6, 7] is different to that used in this work. Since the appropriate space group with relevant cell dimensions is identified for each structure determination (e.g. Tables 1 and 2 of reference[7]), specific details for β–FeSe$_{1-x}$ (listed as α–FeSe in references[3, 6, 7]) are self-evident. Based on general cell parameter trends for orthorhombic β–FeSe$_{1-x}$ (Figure S4b) and derived parameters (e.g. *b/a* and *c/a* ratios) from theses datasets,[7] the modelled cell dimensions at 8 K and 18 GPa may be an anomaly and, while included, are not used for exemplification.

Publications on band structure models for β–FeSe$_{1-x}$ with change in pressure are predominantly focused on the tetragonal P4/nmm space group.[7, 30, 79] Few models accommodate the fundamental phase transformation to a Cmma space group at temperatures < 70 K. As shown in Figure S1, these two structures are different albeit related by a symmetry transformation that has direct bearing on band structure details. Other studies[30, 79] utilize a P1 lattice for computations (as in this study) but notably, the P1 cell derived from a P4/nmm parent symmetry is different to that based on Cmma symmetry evident in values for the interplanar angle, γ. Figures S5 and S6 show that general band topologies of tetragonal and orthorhombic β–FeSe$_{1-x}$ have some similar features. However, details along Γ–Z and at meV scale are not the same.

Acknowledgements

The authors acknowledge the e-Research Office at QUT for access to high performance computing. During the preparation of this manuscript and study, the authors did not use any artificial intelligence tools.

**Data Availability Statement**

Additional data presented in this study are available in Supporting Information. Raw data files from DFT calculations are available from the authors on request.

Received: ((will be filled in by the editorial staff))
Revised: ((will be filled in by the editorial staff))
Published online: ((will be filled in by the editorial staff))

**Supporting Information**